\documentclass[conference]{IEEEtran}

\usepackage[T1]{fontenc}
\usepackage[utf8]{inputenc}
\usepackage[cmex10]{amsmath}
\usepackage{nccmath}
\usepackage{amssymb,amsthm,lipsum}
\usepackage{mathtools}
\usepackage{cuted}
\usepackage{graphicx}
\usepackage{xcolor}
\usepackage{psfrag}
\usepackage{comment}
\usepackage{cite}
\usepackage{balance}
\usepackage{color}
\usepackage{algorithm} 
\usepackage{algpseudocode}
\usepackage{algorithmicx}
\usepackage{cuted}
\usepackage{lipsum}

\theoremstyle{remark}

\DeclareUnicodeCharacter{2212}{-}

\renewcommand{\vec}[1]{{\boldsymbol{\mathrm{#1}}}} 

\begin{document}

\title{Secrecy Capacity Maximization for a Hybrid Relay-RIS Scheme in mmWave MIMO Networks }

\author{\IEEEauthorblockN{Edson Nobuyuki Egashira$^*$, Diana Pamela Moya Osorio$^*$, Nhan Thanh Nguyen$^*$, and Markku Juntti$^*$}
\IEEEauthorblockA{\textit{*Centre for Wireless Communications, University of Oulu, P.O.Box 4500, FI-90014, Finland} \\
Emails: \{edson.egashira, diana.moyaosorio, nhan.nguyen, markku.juntti\}@oulu.fi}}


\maketitle

\begin{abstract}
The hybrid relay-reflecting intelligent surface
(HR-RIS) has been recently introduced as an efficient solution to overcome the double path loss and limited beamforming diversity of the conventional fully passive reflecting surface. This motivates us to investigate the application of the HR-RIS in improving the secrecy capacity of millimeter wave multiple-input-multiple-output (MIMO) systems with the presence of multi-antenna eavesdropper. The joint optimization of the transmit beamformer and the relay-reflecting coefficients at RIS is tackled via alternating optimization. In the proposed solution, a closed-form expression for the optimal transmit beamformer at the transmitter is derived, and a metaheuristic solution based on particle swarm optimization is proposed to optimize the active and passive elements at the HR-RIS. The simulation results verify that under various scenarios, the HR-RIS provides significant improvement in the secrecy capacity with respect to the conventional passive reflecting surface.

\end{abstract}



\section{Introduction}

For the sixth-generation (6G) of wireless networks, it is expected that the radio environment becomes controllable and intelligent by leveraging the emergence of new enabling technologies that allows the control of the wireless propagation in order to improve reliability and coverage. Particularly, reflecting intelligent surfaces~(RIS) have shown a great potential to tackle the challenges of security, energy and spectral efficiency in 6G. RIS is a software-controlled surface composed by a planar array of reflecting elements, which are capable to dynamically adjust their reflective coefficients, thus controlling the amplitude and/or phase shift of reflected signals to enhance the wireless propagation performance~\cite{art:AN_helpful_or_not}. 

By smartly controlling the phase shifts of RIS, the reflected signals can either be added coherently at the intended receiver to improve the received signal power, or be added destructively at the non-desired receiver, thus providing information-theoretic security guarantees and being an important candidate for the design of physical layer security (PLS) schemes~\cite{chap:Osorio2019,art:Joint_active_passive_beamforming}. In this regard, RIS-based PLS schemes have recently attracted increasing attention~\cite{art:MISO_Cs_IRS,art:feng2021,art:RobustandSecure}. For instance, in~\cite{art:MISO_Cs_IRS,art:feng2021}, the secrecy capacity is evaluated for a multiple-input-single-output (MISO) RIS-aided secure system, under the presence of an eavesdropper. In~\cite{art:MISO_Cs_IRS}, Cui {\it et al.} have addressed the impact of the transmit power at the access point (AP) on the secrecy capacity by using a joint alternating optimization (AO)- based and semidefinite relaxation (SDR) methods. In~\cite{art:feng2021}, a fractional programming (FP) and Manifold Optimization (MO)- based algorithm is proposed to maximize the secrecy capacity.
In~\cite{art:RobustandSecure}, the average system sum secrecy rate and secrecy outage probability were analyzed for a system with multiple users, eavesdroppers and RIS. By taking into account the imperfect CSI, the results confirm the robustness and the ability to exploit the multiuser diversity.  

Particularly, in the context of millimeter-wave (mmWave) multiple-input-multiple-output (MIMO) systems that are recognized for providing ultra-high data-rates, RIS can offer an important solution to the  high propagation loss nature of these systems in order to improve reliability and security~\cite{art:SecureTransmissionmmWave,art:xiu2021}. Thus, the authors in~\cite{art:SecureTransmissionmmWave} investigate the secrecy rate of a RIS-assisted mmWave MISO system in the presence of a single-antenna eavesdropper. Therein, an Alternating Optimization (AO) approach is proposed to jointly optimize the transmit beamformer at the base station and the passive reflecting elements at the RIS. The proposed method has shown to boost the secrecy performance, regardless of the eavesdropper's positions. In~\cite{art:xiu2021}, an AO-based algorithm is proposed to maximize the secrecy rate for a mmWave MISO system by assuming multiple RISs and a single-antenna passive eavesdropper. The spatially distributed multiple RISs deployment showed a significant secrecy improvement compared with a single RIS.    

Despite the conventional RIS has shown to be a promising technology to enable secure and intelligent radio environments, it has been proved that a very large number of reflecting elements of a passive RIS is needed to outperform the performance of decode-and-forward relaying~\cite{bjornson_intelligent_2019}. Therefore, hybrid relay-RIS (HR-RIS) scheme was introduced in~\cite{art:nhan2022hybrid} to leverage the benefits from both passive RIS and relay. Specifically, Nguyen \textit{et al.} proposed employing a single or few active elements in the RIS to provide the signal amplification gain for a significant improvement in the system spectral efficiency. 


Inspired by the promising simultaneous reflecting-relaying capabilities of the HR-RIS, this paper advances on the state-of-the art by investigating and optimizing the secrecy performance of the HR-RIS scheme in a mmWave MIMO system with the presence of a multiple-antenna eavesdropper. For this purpose, the joint optimization of the transmit beamformer and the relay-reflecting coefficients at RIS is attained by resorting to AO. Specifically, the closed-form solution to the optimal transmit beamformer is first derived with given HR-RIS coefficients. Then, we employ the particle swarm optimization (PSO) approach to tackle the challenges in the HR-RIS coefficients design. Our simulations for various scenarios show that the HR-RIS-aided mmWave MIMO system significantly outperforms its conventional counterparts, including the systems without RIS and with the passive RIS.

\section{System Model}
\begin{figure}[htp]
    \centering
    \includegraphics[width=8cm]{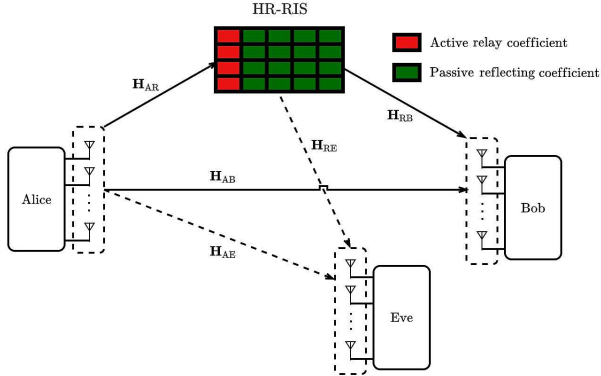}
    \caption{An HR-RIS assisted mmWave MIMO system, with muiti-antenna Alice, multi-antenna Bob, HR-RIS, and a multi-antenna Eve.}
    \label{fig:system_model_conference}
\end{figure}
We consider a mmWave MIMO system consisting of a transmitter ($\mathrm{Alice}$), a receiver ($\mathrm{Bob}$), and an eavesdropper (Eve). The eavesdropper is an active user, and the communications system is assisted by an HR-RIS, as shown in Fig.~\ref{fig:system_model_conference}.
Let $N_\mathrm{A}, N_\mathrm{B}$ and $N_\mathrm{E}$ be the numbers of antennas at $\mathrm{Alice}$, $\mathrm{Bob}$ and Eve,  respectively, and the HR-RIS is equipped with $N$ elements, comprising $K$ active relay elements and $M$ passive reflecting elements, such that $K + M = N$. 

\subsection{Channel Model}

Let us denote the channel matrices between $\mathrm{Alice}$ and HR-RIS, between HR-RIS and $\mathrm{Bob}$, between HR-RIS and $\mathrm{Eve}$, between $\mathrm{Alice} $ and $\mathrm{Bob}$, and between $\mathrm{Alice}$ and $\mathrm{Eve}$ as ${\vec H}_\mathrm{AR} \in \mathbb{C}^{N \times N_\mathrm{A}}$, ${\vec H}_\mathrm{RB} \in \mathbb{C}^{N_\mathrm{B} \times N}$, ${\vec H}_\mathrm{RE} \in \mathbb{C}^{N_\mathrm{E} \times N}$, ${\vec H}_\mathrm{AB} \in \mathbb{C}^{N_\mathrm{B} \times N_\mathrm{A}}$, and ${\vec H}_\mathrm{AE} \in \mathbb{C}^{N_\mathrm{E} \times N_\mathrm{A}}$, respectively.
We assume that $\mathrm{Alice}$ is able to acquire the CSI of the $\mathrm{Bob}$ perfectly.
However, the eavesdropper node may try to hide its existence from $\mathrm{Alice}$ and despite the signal leakage from $\mathrm{Eve}$ can still be utilized for channel estimation at $\mathrm{Alice}$, the acquired CSI is expected to be outdated~\cite{art:RobustandSecure}. Hence, we adopt the imperfect CSI model, where the channel coefficients between HR-RIS and $\mathrm{Eve}$ and between $\mathrm{Alice}$ and $\mathrm{Eve}$ are modeled by~${\vec H}_\mathrm{X} = \bar{{\vec H}}_\mathrm{X} +  \boldsymbol{\Delta \mathrm{H}}_\mathrm{X}$, 
where $\mathrm{X} = \{\mathrm{RE},\mathrm{AE}\}$, $\bar{{\vec H}}_\mathrm{X}$ is the outdated estimate of the eavesdropper's channel, and $\boldsymbol{\Delta \mathrm{H}}_\mathrm{X}$ is the CSI estimation error, modeled with the Gaussian CSI error model~\cite{art:uncertainty1mmWave}. Thus, the CSI estimation error $\boldsymbol{\mathrm{\Delta H}}_\mathrm{RE}$ is given by
\begin{align}
    \boldsymbol{\mathrm{\Delta H}}_\mathrm{RE} &= \sqrt{\frac{N N_\mathrm{E}}{L_\mathrm{RE}}}\boldsymbol{\delta}_\mathrm{RE}
    \Big(\beta_\mathrm{RE_1} \boldsymbol{\mathrm{a}}(\phi_\mathrm{RE}^r)\boldsymbol{\mathrm{a}}^\mathrm{H}(\gamma_\mathrm{RE}^t, \eta_\mathrm{RE}^t) \nonumber\\
    &\quad+ \sum_{i=2}^{L_\mathrm{RE}} \beta_\mathrm{RE_i} \boldsymbol{\mathrm{a}}(\phi_\mathrm{RE}^r)\boldsymbol{\mathrm{a}}^\mathrm{H}(\gamma_\mathrm{RE}^t, \eta_\mathrm{RE}^t)\Big),\label{eq:DeltaHRE}
\end{align}
where $\boldsymbol{\delta}_\mathrm{RE} \sim \mathcal{CN}(0, {\vec I}_{N_\mathrm{E}} \sigma_{\Delta}^2)$, with $\mathcal{CN}(a,\,b)$ denoting as a complex circularly-symmetric Gaussian distribution with mean $a$ and variance $b$, ${\vec I}_{N}$ denotes an identity matrix of size $N \times N$, and $\sigma_{\Delta}^2$ is the normalized variance of the error component of the channel. $L_\mathrm{RE}$ is the number of resolvable paths, $\beta_\mathrm{{RE}_1}$ and $\beta_\mathrm{{RE}_i}$ are the complex gain, $\boldsymbol{\mathrm{a}}(\phi_\mathrm{RE}^r)$ and $\boldsymbol{\mathrm{a}}(\gamma_\mathrm{RE}^t, \eta_\mathrm{RE}^t)$ are the normalized array response vectors. For the channel $\boldsymbol{\mathrm{\Delta H}}_\mathrm{AE}$, a similar procedure is applied as in~\eqref{eq:DeltaHRE}.

In addition, all the channels are assumed to follow the Saleh-Valenzuela channel model \cite{initialSaleh}, which is widely assumed for mmWave systems~\cite{art:SecureTransmissionmmWave}. As a result, by considering uniform linear arrays (ULA) at the Alice, Bob, and Eve's nodes, and a uniform planar array (UPA) for the HR-RIS, the Alice$\rightarrow$ HR-RIS channel is given by
\begin{align}
    {\vec H}_\mathrm{AR} &= \sqrt{\frac{N_\mathrm{A} N}{L_\mathrm{AR}}} \Big( \mathrm{\beta}_\mathrm{AR1} \boldsymbol{\mathrm{a}}(\gamma_\mathrm{AR}^r, \eta_\mathrm{AR}^r)\boldsymbol{\mathrm{a}}^\mathrm{H}(\phi_\mathrm{AR}^t)\nonumber \\
    &+ \sum_{i=2}^{L_\mathrm{AR}} \beta_{\mathrm{AR}i} \boldsymbol{\mathrm{a}}(\gamma_\mathrm{AR}^r, \eta_\mathrm{AR}^r)\boldsymbol{\mathrm{a}}^\mathrm{H}(\phi_\mathrm{AR}^t)\Big), \label{eq:HAR}
\end{align}
where $L_\mathrm{AR}$ is the number of resolvable paths, $\beta_\mathrm{{AR}1}\sim \mathcal{CN}(0, 10^{−0.1\kappa_\mathrm{{AR}1}})$ is the complex gain of the line-of-sight (LOS) path, $\beta_{\mathrm{AR}i} \sim \mathcal{CN}(0, 10^{−0.1\kappa_{\mathrm{AR}i}})$ is the complex gain of the $i$-th non-LOS (NLOS) path. The path losses $\kappa_{\mathrm{AR}1}$ and $\kappa_{\mathrm{AR}i}$ are given~by
\begin{align}
    \kappa_{\mathrm{AR}1} &= a_\mathrm{LOS} + 10 b_\mathrm{LOS} \log_{10}(d_\mathrm{AR}) + \mu_\mathrm{LOS}, \label{eq:kappa1}\\
    \kappa_{\mathrm{AR}i} &= a_\mathrm{NLOS} + 10 b_\mathrm{NLOS} \log_{10}(d_\mathrm{AR}) + \mu_\mathrm{NLOS}, \label{eq:kappai}
\end{align}
where $d_\mathrm{AR}$ is the distance between Alice and HR-RIS, and $\{\mu_\mathrm{LOS},~\mu_\mathrm{NLOS}\} \sim \{\mathcal{N}(0, \sigma^2_{\mu_\mathrm{LOS}}), \mathcal{N}(0,~ \sigma^2_{\mu_\mathrm{NLOS}})\}$, where $\mathcal{N}(a,\,b)$ denotes a Gaussian distribution with mean $a$ and variance $b$. From~\eqref{eq:HAR}, $\boldsymbol{\mathrm{a}}(\gamma_\mathrm{AR}^r, \eta_\mathrm{AR}^r)$ and $\boldsymbol{\mathrm{a}}(\phi_\mathrm{AR}^t)$ are the normalized array response vectors associated with the receiver and transmitter node, respectively, $\gamma_\mathrm{AR}^r$ and $\eta_\mathrm{AR}^r$ are their azimuth and elevation angles of arrival (AoA) associated with the HR-RIS, and $\phi_\mathrm{AR}^t$ is the angle of departure (AoD) associated with the Alice node. Note that $\boldsymbol{\mathrm{a}}^\mathrm{H}$ denotes the Hermitian transpose of a vector $\boldsymbol{\mathrm{a}}$. By assuming an antenna spacing of $\lambda/2$, the normalized array response vectors can be expressed as
\begin{align}
[\boldsymbol{\mathrm{a}}(\gamma_\mathrm{AR}^r, \eta_\mathrm{AR}^r)]_{n} &= \frac{1}{\sqrt{N}} [e^{j\pi (m-1) \cos(\eta_\mathrm{AR}^r)\sin(\gamma_\mathrm{AR}^r)}  \nonumber\\
&\quad \times e^{j\pi(n-1)\sin(\eta_\mathrm{AR}^r)}]^\mathrm{T}, m = 1,\cdots, N_x \nonumber\\
&\quad n = 1,\cdots, N_y, \\
    [\boldsymbol{\mathrm{a}}(\phi_\mathrm{AR}^t)]_k &= e^{j\pi (k-1) \sin(\theta_{Xi})}, k = 1,\cdots, N_\mathrm{A}, \label{eq:responsevecAR}
\end{align}
where $N_x$ and $N_y$ are the number of rows and columns of coefficients for a given HR-RIS, such that $N_x \times N_y = N$. Additionally, the remaining channels ${\vec H}_\mathrm{RB}, {\vec H}_\mathrm{AB}, \bar{{\vec H}}_\mathrm{RE}$, and $\bar{{\vec H}}_\mathrm{AE}$ can be modeled similarly as in~\eqref{eq:HAR}--\eqref{eq:responsevecAR}, and more details can be found in~\cite{art:MIMO_mmWave_LIS}.

\subsection{Signal Model}
For the optimal design of reflecting/amplifying coefficients, the phases and amplitudes of the incident signals are adjusted so that the signals are added constructively at $\mathrm{Bob}$. Let
\begin{equation} \label{eq:alphan}
\alpha_n = 
    \begin{cases}
      |\alpha_n| e^{j \theta_n}, & \text{if}~n\in \mathbb{A}\\
      e^{j \theta_n},  & \text{otherwise}
    \end{cases}
\end{equation}
denote the reflecting/amplifying coefficient associated with the $n$th element of the HR-RIS, where $\theta_n \in ( 0, 2\pi ]$ is the phase shift and $\mathbb{A}$ is the set of the positions of the active elements. Accordingly, let $\boldsymbol{\Upsilon} =$ diag$\{\alpha_1, ..., \alpha_N\} \in \mathbb{C}^{N \times N}$ be the diagonal matrix of the HR-RIS's coefficients, and $\boldsymbol{\Psi} \in \mathbb{C}^{N \times N}$ be the diagonal matrix including only the active elements of HR-RIS. 
Therefore, the received signal at $\mathrm{Bob}$, denoted by $\boldsymbol{y}_\mathrm{B}\in \mathbb{C}^{N_\mathrm{B}\times 1}$, can be expressed as
\begin{align}
    \boldsymbol{y}_\mathrm{B} &= ({\vec H}_\mathrm{RB} \boldsymbol{\Upsilon} \mathrm{{\vec H}_\mathrm{AR}} + {\vec H}_\mathrm{AB})\boldsymbol{s}  + {\vec H}_\mathrm{RB} \boldsymbol{\Psi} \boldsymbol{n}_\mathrm{R} + \boldsymbol{n}_\mathrm{B} \nonumber\\
    &= ({\vec H}_\mathrm{RB} \boldsymbol{\Upsilon} \mathrm{{\vec H}_\mathrm{AR}} + {\vec H}_\mathrm{AB}) \boldsymbol{s}   + \boldsymbol{n}_\mathrm{T},
\end{align}
where 
$\boldsymbol{n}_\mathrm{R} \in \mathbb{C}^{N\times 1} $ and $\boldsymbol{n}_\mathrm{B}\in \mathbb{C}^{N_\mathrm{B}\times 1}$ are the complex additive Gaussian noise (AWGN) modeled with zero-mean circularly symmetric complex Gaussian, with covariance matrices $\sigma_\mathrm{H}^2 {\vec I}_N$ and $\sigma_\mathrm{B}^2 {\vec I}_{N_\mathrm{B}}$, respectively. Note that $\boldsymbol{n}_\mathrm{T} = {\vec H}_\mathrm{RB} \boldsymbol{\Psi} \boldsymbol{n}_\mathrm{R} + \boldsymbol{n}_\mathrm{B}$ is the total effective noise at $\mathrm{Bob}$. Similarly, the received signal at Eve, i.e.,~$\boldsymbol{\mathrm{y}}_\mathrm{E}\in \mathbb{C}^{N_\mathrm{E}\times 1}$ is given by
\begin{align}
    \boldsymbol{\mathrm{y}}_\mathrm{E} &= ({\vec H}_\mathrm{RE} \boldsymbol{\Upsilon} {\vec H}_\mathrm{AR} + {\vec H}_\mathrm{AE}) \boldsymbol{s}  + {\vec H}_\mathrm{RE} \boldsymbol{\Psi} \boldsymbol{n}_\mathrm{R} + \boldsymbol{n}_\mathrm{E} \nonumber\\
    &= ({\vec H}_\mathrm{RE} \boldsymbol{\Upsilon} {\vec H}_\mathrm{AR}+ {\vec H}_\mathrm{AE}) \boldsymbol{s}   + \boldsymbol{n}_\mathrm{TE},
\end{align}
where $\boldsymbol{n}_\mathrm{E}\in \mathbb{C}^{N_\mathrm{E}\times 1}$ is the AWGN with covariance matrix $\sigma_\mathrm{E}^2 {\vec I}_{N_\mathrm{E}}$, and $\boldsymbol{n}_\mathrm{TE} = {\vec H}_\mathrm{RB} \boldsymbol{\Psi} \boldsymbol{n}_\mathrm{R} + \boldsymbol{n}_\mathrm{E}$. For ease of exposition, we assume that $\sigma_\mathrm{H}^2 = \sigma_\mathrm{B}^2 = \sigma_\mathrm{E}^2 = \sigma^2$, $\boldsymbol{n}_\mathrm{T}\sim \mathcal{CN}(0,\,\sigma^{2}({\vec I}_{N_\mathrm{B}} + {\vec H}_\mathrm{RB} \boldsymbol{\Psi}\boldsymbol{\Psi}^\mathrm{H} {\vec H}_\mathrm{RB}^\mathrm{H}))$ and $\boldsymbol{n}_\mathrm{TE}\sim \mathcal{CN}(0,\,\sigma^{2}({\vec I}_{N_\mathrm{E}} + {\vec H}_\mathrm{RE} \boldsymbol{\Psi}\boldsymbol{\Psi}^\mathrm{H} {\vec H}_\mathrm{RE}^\mathrm{H}))$. For simplicity, let us consider ${\vec H}_\mathrm{B} = {\vec H}_\mathrm{RB} \boldsymbol{\Upsilon} {\vec H}_\mathrm{AR} + {\vec H}_\mathrm{AB}$ and ${\vec H}_\mathrm{E} = {\vec H}_\mathrm{RE} \boldsymbol{\Upsilon} {\vec H}_\mathrm{AR}+ {\vec H}_\mathrm{AE}$. Hence, the channel capacities between Alice and Bob and between Alice and Eve can be respectively given by
\begin{align} 
    C_{l} &= \log_2 \bigg(\det\bigg( {\vec I}_{N_\mathrm{B}} + \frac{1}{\sigma^2}{\vec Q}_\mathrm{B}^{-1}\vec H_\mathrm{B} {\vec Q}_\mathrm{A} {\vec H}_\mathrm{B}^\mathrm{H} \bigg)\bigg) \label{eq:C_B} \\
    C_{e} &= \log_2 \bigg(\det\bigg( {\vec I}_{N_\mathrm{E}} + \frac{1}{\sigma^2}{\vec Q}_\mathrm{E}^{-1}{\vec H}_\mathrm{E} {\vec Q}_\mathrm{A} {\vec H}_\mathrm{E}^\mathrm{H} \bigg)\bigg), \label{eq:C_E}
\end{align}
respectively, where ${\vec Q}_\mathrm{A} \in \mathbb{C}^{N_\mathrm{A} \times N_\mathrm{A}} = \mathbb{E}\{\boldsymbol{s} \boldsymbol{s}^\mathrm{H}\}$ is the transmit covariance matrix, satisfying $\mathrm{tr}({\vec Q}_\mathrm{A}) \leq P_T$, in which $P_T$ is the transmit power at Alice, and $\mathrm{tr}(\cdot)$ denotes the trace of a squared matrix; ${\vec Q}_\mathrm{B} \in \mathbb{C}^{N_\mathrm{B} \times N_\mathrm{B}} = {\vec I}_{N_\mathrm{B}} + {\vec H}_\mathrm{RB} \boldsymbol{\Psi}\boldsymbol{\Psi}^\mathrm{H} {\vec H}_\mathrm{RB}^\mathrm{H}$ and ${\vec Q}_\mathrm{E} \in \mathbb{C}^{N_\mathrm{E} \times N_\mathrm{E}} = {\vec I}_{N_\mathrm{E}} + {\vec H}_\mathrm{RE} \boldsymbol{\Psi}\boldsymbol{\Psi}^\mathrm{H} {\vec H}_\mathrm{RE}^\mathrm{H}$ are the aggregate noise covariance matrices. By considering a single-stream beamforming scheme~\cite{art:wang2021}, the optimal solution to ${\vec Q}_\mathrm{A}$ has rank-one and can be expressed as ${\vec Q}_\mathrm{A} = \boldsymbol{\mathrm{w}} \boldsymbol{\mathrm{w}}^\mathrm{H}$, where $\boldsymbol{\mathrm{w}} \in \mathbb{C}^{N_\mathrm{A} \times 1}$ is the optimal transmit beamforming vector. Thus, by replacing the optimal ${\vec Q}_\mathrm{A}$ into~\eqref{eq:C_B} and~\eqref{eq:C_E}, the channel capacities can be expressed as 
\begin{align} 
    C_{l} &= \log_2 \bigg(\det\bigg( {\vec I}_{N_\mathrm{B}} +  \frac{1}{\sigma^2}{\vec Q}_\mathrm{B}^{-1}{\vec H}_\mathrm{B} {\vec w} {\vec w} ^\mathrm{H} {\vec H}_\mathrm{B}^\mathrm{H} \bigg)\bigg) \label{eq:C_B2} \\
    C_{e} &= \log_2 \bigg(\det\bigg( {\vec I}_{N_\mathrm{E}} + \frac{1}{\sigma^2}{\vec Q}_\mathrm{E}^{-1}{\vec H}_\mathrm{E} {\vec w} {\vec w} ^\mathrm{H} {\vec H}_\mathrm{E}^\mathrm{H} \bigg)\bigg), \label{eq:C_E2}
\end{align}
From~\eqref{eq:C_B2} and~\eqref{eq:C_E2}, the secrecy capacity can be obtained as
\begin{align} 
    C_{S} &= [C_{l} - C_{e}]^{+} \triangleq \mathrm{max}[C_{l} - C_{e}, 0]. \label{eq:Cs}
\end{align}

\section{Secrecy Capacity Maximization}

In this section, the solution to the secrecy capacity maximization problem is obtained via an alternating optimization (AO), which alternatively optimize the transmit beamforming at  Alice, and the HR-RIS coefficients. More specifically, we first find the solution to ${\vec w} $ for a given $\alpha_{n}$, and then, the optimal solution to $\alpha_{n}$ are obtained for a given ${\vec w} $.

\subsection{Transmit Beamforming Optimization}
First, with fixed HR-RIS coefficients, i.e., fixed $\alpha_{n}$, we aim at finding the optimal ${\vec w} $ by solving the following problem:
\begin{equation}
\begin{aligned}
\max_{{\vec w} } \quad & C_{l} - C_{e}\\
\textrm{s.t.} \quad & ||{\vec w} ||^2 \leq P_T,
\end{aligned} \label{opt:Cs}
\end{equation}
Note that we have neglected the operator $[\cdot]^{+}$ from~\eqref{eq:Cs} due to the fact that the optimal values must be non-negative~\cite{art:MISO_Cs_IRS}.

According to the Sylvester's determinant identity, we have that $\det ({\vec I}_P + \boldsymbol{\mathrm{AB}}) = \det ({\vec I}_Q + \boldsymbol{\mathrm{BA}})$ for arbitrary matrices  $\boldsymbol{\mathrm{A}}\in \mathbb{C}^{P \times Q}$ and $\boldsymbol{\mathrm{B}}\in \mathbb{C}^{Q \times P}$. Therefore,  $C_l$ and $C_e$ in ~\eqref{eq:C_B2} and~\eqref{eq:C_E2} can be rewritten as 
\begin{align} 
    C_{l} &= 
    \log_2 \big(\det\big( {\vec I}_{N_\mathrm{B}} + \frac{1}{\sigma^2}{\vec Q}_\mathrm{B}^{-1}{\vec H}_\mathrm{B} {\vec w} {\vec w}^\mathrm{H}  {\vec H}_\mathrm{B}^\mathrm{H}  \big)\big) \nonumber \\
    &\quad = 
    \log_2 \big( 1 + \frac{1}{\sigma^2} {\vec w}^\mathrm{H}  {\vec H}_\mathrm{B}^\mathrm{H} {\vec Q}_\mathrm{B}^{-1}{\vec H}_\mathrm{B} {\vec w}   \big), \label{eq:C_B_nodet} \\
      C_{e} &= 
      \log_2 \big(\det\big( {\vec I}_{N_\mathrm{E}} + \frac{1}{\sigma^2} {\vec Q}_\mathrm{E}^{-1}{\vec H}_\mathrm{E} {\vec w} {\vec w}^\mathrm{H}  {\vec H}_\mathrm{E}^\mathrm{H} \big)\big) \nonumber \\
    &\quad = 
    \log_2 \big( 1 + \frac{1}{\sigma^2} {\vec w}^\mathrm{H}  {\vec H}_\mathrm{E}^\mathrm{H} {\vec Q}_\mathrm{E}^{-1}{\vec H}_\mathrm{E} {\vec w}  \big). \label{eq:C_E_nodet}
\end{align}
Thus, problem \eqref{opt:Cs} can be re-expressed as
\begin{equation}
\begin{aligned}
\max_{{\vec w} } \quad & \log_2 \Bigg( \frac{ 1 + \frac{1}{\sigma^2} 
{\vec w}^\mathrm{H}  {\vec H}_\mathrm{B}^\mathrm{H} {\vec Q}_\mathrm{B}^{-1}{\vec H}_\mathrm{B} {\vec w} }
{ 1 + \frac{1}{\sigma^2}  {\vec w}^\mathrm{H}  {\vec H}_\mathrm{E}^\mathrm{H} {\vec Q}_\mathrm{E}^{-1}{\vec H}_\mathrm{E} {\vec w} } \Bigg) \\
\textrm{s.t.} \quad & ||{\vec w} ||^2 \leq P_T.  \\
\end{aligned} \label{opt:Cs_phase1}
\end{equation}
For ease of exposition, let us denote $\boldsymbol{\mathrm{L}} = \frac{1}{\sigma^2}  {\vec H}_\mathrm{B}^\mathrm{H} {\vec Q}_\mathrm{B}^{-1}{\vec H}_\mathrm{B}$, and $\boldsymbol{\mathrm{E}} = \frac{1}{\sigma^2}  {\vec H}_\mathrm{E}^\mathrm{H} {\vec Q}_\mathrm{E}^{-1}{\vec H}_\mathrm{E}$, which are constant with respect to variable ${\vec w} $. Furthermore, due to the monotonic increasing property of $\log_2(\cdot)$, we can rewrite \eqref{opt:Cs_phase1} as
\begin{equation}
\begin{aligned}
\max_{{\vec w} } \quad & \frac{1 + {\vec w}^\mathrm{H}  \boldsymbol{\mathrm{L}} {\vec w} }{1 + {\vec w}^\mathrm{H}  \boldsymbol{\mathrm{E}} {\vec w} }\\
\textrm{s.t.} \quad & ||{\vec w} ||^2 \leq P_T,  \\
\end{aligned} \label{form:beamforming}
\end{equation}
Finally, the optimal solution to $\boldsymbol{\mathrm{w}}$ can be found as~\cite{art:MISOME_IRS_det_sum}
\begin{equation}
    \boldsymbol{\mathrm{w}^*} = \sqrt{P_T}{\vec v}_\mathrm{m}\bigg[\bigg(\boldsymbol{\mathrm{E}} + \frac{1}{P_T}{\vec I}_{N_\mathrm{A}}\bigg)^{-1}\bigg(\boldsymbol{\mathrm{L}} + \frac{1}{P_T}{\vec I}_{N_\mathrm{A}}\bigg)\bigg], \label{eq:wOP}
\end{equation}
where ${\vec v}_\mathrm{m}[\cdot]$ represents the normalized eigenvector corresponding to the largest eigenvalue of its argument.

\subsection{Optimization of the Reflecting Coefficients of the HR-RIS}
With given transmit beamforming vector $\boldsymbol{\mathrm{w}}$, the optimal solution to the HR-RIS coefficients can be found by solving the following problem:
\begin{equation}
\begin{aligned}
\max_{|\alpha_{n}|,\theta_{n}} \quad & \log_2 \Bigg(\frac{1 + {\vec w}^\mathrm{H}  \boldsymbol{\mathrm{L}} {\vec w} }{1 + {\vec w}^\mathrm{H}  \boldsymbol{\mathrm{E}} {\vec w} }\Bigg)\\
\textrm{s.t.} \quad & P_a(\alpha_{n}) \leq P_\mathrm{max}   \\
  &|\alpha_{n}| = 1,~\textrm{for}~n \notin \mathbb{A},  
\end{aligned} \label{opt:Cs2}
\end{equation}
where  $P_\mathrm{max}$ is the power budget of the active elements, and $P_a(\alpha_{n})$ is the transmit power of active elements at HR-RIS, given by
\begin{align}
    \nonumber P_a(\alpha_{n}) =& \text{tr}(\boldsymbol{\Psi} ( {\vec H}_\mathrm{AR}{\vec Q}_\mathrm{A}{\vec H}_\mathrm{AR}^H + \sigma^2 {\vec I}_N) \boldsymbol{\Psi}^\mathrm{H})  \\
    \nonumber=& \text{tr}(\boldsymbol{\Psi}\Tilde{\vec H}_\mathrm{AR}\Tilde{\vec H}_\mathrm{AR}^H\boldsymbol{\Psi}) + \sigma^2\text{tr}(\boldsymbol{\Psi}\boldsymbol{\Psi}^H) \\
     =& \sum_{n \in \mathbb{A}} (|\alpha_{n}|^2 ||\Tilde{\vec{t}_n}||^2) + \sigma^2\sum_{n \in \mathbb{A}} |\alpha_{n}|^2 \nonumber \\
     =& \sum_{n \in \mathbb{A}} |\alpha_{n}|^2 \xi_n,\label{eq:Pamaxconstraint}
\end{align}
where $\xi_n \triangleq |\Tilde{t}_n|^2 + \sigma^2$, and $\Tilde{t}_n$ is the ${n}$th element of $\Tilde{\vec h}_\mathrm{AR} \triangleq {\vec H}_\mathrm{AR}{\vec w}  \in \mathbb{C}^{N \times 1}$. On the other hand, let  $\vec{r}_n  \in \mathbb{C}^{N_\mathrm{B} \times 1}$ be the $n$th column of ${\vec H}_\mathrm{RB}$, $\vec{t}_n^\mathrm{H}  \in \mathbb{C}^{1 \times N_\mathrm{A}}$ be the $n$th row of ${\vec H}_\mathrm{AR}$, and $\vec{e}_n  \in \mathbb{C}^{N_\mathrm{E} \times 1}$ be the $n$th column of ${\vec H}_\mathrm{RE}$ such that ${\vec H}_\mathrm{AR} = \left[\vec{t}_1, ..., \vec{t}_N \right]^\mathrm{H}$, ${\vec H}_\mathrm{RB} = \left[\vec{r}_1, ..., \vec{r}_N \right]$ and ${\vec H}_\mathrm{RE} = \left[\vec{e}_1, ..., \vec{e}_N \right]$, respectively. By noticing that $\boldsymbol{\Upsilon}$ and $\boldsymbol{\Psi}$ are diagonal matrices, we can express the effective channels as 
\begin{align}
    {\vec H}_\mathrm{B} = \sum_{n = 1}^{N} \alpha_n \boldsymbol{r}_n \boldsymbol{t}_n^\mathrm{H} + {\vec H}_\mathrm{AB} \label{eq:HBnew}\\
    {\vec H}_\mathrm{E} = \sum_{n = 1}^{N} \alpha_n \boldsymbol{e}_n \boldsymbol{t}_n^\mathrm{H}+ {\vec H}_\mathrm{AE}, \label{eq:HEnew}
\end{align}
and similarly, the covariance matrices can be reformulated as
\begin{align}
    {\vec Q}_\mathrm{B} = {\vec I}_{N_\mathrm{B}} {+} \Bigg( \sum_{n \in \mathbb{A}} \alpha_n \boldsymbol{r}_n \Bigg) \Bigg( \sum_{n \in \mathbb{A}} \alpha_n \boldsymbol{r}_n \Bigg)^\mathrm{H}\label{eq:QBnew}\\
     {\vec Q}_\mathrm{E} = {\vec I}_{N_\mathrm{E}} + \Bigg( \sum_{n \in \mathbb{A}} \alpha_n \boldsymbol{e}_n \Bigg) \Bigg( \sum_{n \in \mathbb{A}} \alpha_n \boldsymbol{e}_n \Bigg)^\mathrm{H}.\label{eq:QEnew}
\end{align}
Thus, from~(\ref{eq:Pamaxconstraint})--(\ref{eq:QEnew}) we can reformulate~\eqref{opt:Cs2} as in~\eqref{opt:Cs_explicit}, shown on the top of the next page. Note we can also express the optimization problem in terms of ${|\alpha_{n}|}$ and $\theta_{n}$ by replacing~\eqref{eq:alphan} into~\eqref{opt:Cs_explicit}.
\begin{figure*}
\begin{equation}
\begin{aligned}
\max_{\alpha_{n}} \quad & \log_2 \Bigg( \frac{ 1 {+} \frac{1}{\sigma^2}
{\vec w}^\mathrm{H}  (\sum_{n = 1}^{N} \alpha_n \boldsymbol{r}_n \boldsymbol{t}_n^\mathrm{H} {+} {\vec H}_\mathrm{AB})^\mathrm{H} \{[{\vec I}_{N_\mathrm{B}} + ( \sum_{n \in \mathbb{A}} \alpha_n \boldsymbol{r}_n ) ( \sum_{n \in \mathbb{A}} \alpha_n \boldsymbol{r}_n )]^\mathrm{H}\}^{-1}(\sum_{n = 1}^{N} \alpha_n \boldsymbol{r}_n \boldsymbol{t}_n^\mathrm{H} {+} {\vec H}_\mathrm{AB}) {\vec w} }
{ 1 {+} \frac{1}{\sigma^2}
{\vec w}^\mathrm{H}  (\sum_{n = 1}^{N} \alpha_n \boldsymbol{e}_n \boldsymbol{t}_n^\mathrm{H} {+} {\vec H}_\mathrm{AE})^\mathrm{H} \{[{\vec I}_{N_\mathrm{E}} + ( \sum_{n \in \mathbb{A}} \alpha_n \boldsymbol{e}_n ) ( \sum_{n \in \mathbb{A}} \alpha_n \boldsymbol{e}_n )]^\mathrm{H}\}^{-1}(\sum_{n = 1}^{N} \alpha_n \boldsymbol{e}_n \boldsymbol{t}_n^\mathrm{H} {+} {\vec H}_\mathrm{AE}) {\vec w}  } \Bigg) \\
\textrm{s.t.} \quad & \sum_{n \in \mathbb{A}} |\alpha_{n}|^2 \xi_n \leq P_\mathrm{max},\\
 \quad &|\alpha_{n}| = 1,~\textrm{for}~n \notin \mathbb{A},  
\end{aligned} \label{opt:Cs_explicit}
\end{equation}
\vspace{-2cm}
\begin{strip}
\noindent\rule{\textwidth}{0.3pt}
\end{strip}
\end{figure*}

Due to the non-convexity of the optimization problem in~\eqref{opt:Cs_explicit} the optimal values for the active and passive elements of the HR-RIS cannot be obtained in a closed form. However, we can provide local optimum results by applying Particle Swarm Optimization (PSO) method as in~\cite{art:clerc2002particle}. The proposed PSO method is presented in Algorithm 1.
\begin{algorithm}
\caption{PSO-based HR-RIS coefficients optimization}
\label{pseudoPSO}
\begin{algorithmic}[1]
\State Set $m_\mathrm{I}$, $n_\mathrm{P}$, ${\vec x}_\mathrm{min}$, ${\vec x}_\mathrm{max}$, $w$, $c_1$, $c_2$, $CF$;
\State Randomly initialize the particles ;
\For{$t$ = 1 : $m_\mathrm{I}$}
\For{$i$ = 1 : $n_\mathrm{P}$}
    \State Update the velocity: ${\vec v}_i(t +1) = w*{\vec v}_i(t) + r_1*c_1*({\vec p}_i(t)-{\vec x}_i(t)) + r_2*c_2*({\vec p}_\mathrm{g}(t)-{\vec x}_i(t))$;
    \State Update the position: ${\vec x}_i(t+1) = {\vec x}_i(t) + {\vec v}_i(t +1)$;
    \State Evaluate the $CF${$({\vec x}_i(t+1))$};
    \State Update the personal and global best positions: 
    \If{$CF$(${\vec x}_i(t+1)$~<~$CF$(${\vec p}_i(t))$}
       \State ${\vec p}_i(t+1) \gets {\vec x}_i(t+1)$
       \If{$CF$(${\vec p}_i(t+1)$)~<~$CF$(${\vec p}_\mathrm{g}(t))$}
       \State ${\vec p}_\mathrm{g}(t+1) \gets {\vec x}_i(t+1)$
    \EndIf
    \EndIf
\EndFor
\EndFor
\end{algorithmic}
\end{algorithm}
For the initial parameters, the maximum number of iterations, the number of particles for the population, the inertia coefficient, the acceleration coefficient for the cognitive component, the acceleration for the global component, and the cost function are denoted by~$m_\mathrm{I}$, $n_\mathrm{P}$, $w$, $c_1$, $c_2$,~$CF$, respectively. The vectors ${\vec x}_\mathrm{min}, {\vec x}_\mathrm{max}$ represent the range of all reflecting coefficients, where ${\vec x}_\mathrm{min} \in \mathbb{C}^{1 \times 2N}$ = $[1, ..., 1, 0, ..., 0]$, and ${\vec x}_\mathrm{max} \in \mathbb{C}^{1 \times 2N} = [\sqrt{P_\mathrm{max}/\min \{\xi_{1,..., K}\}}, ..., \sqrt{P_\mathrm{max}/\min \{\xi_{1,..., K}\}},...,1, ...,\\1, ..., 2\pi, ..., 2\pi]$. To ensure convergence, the stability and convergence criteria are adopted based on constriction coefficient~\cite{art:clerc2002particle}, given by
\begin{align}
    \chi = \frac{2}{|2-\kappa - \sqrt{\kappa^2 - 4\kappa}|}, \label{eq:contrictioncoef}
\end{align}
where $\kappa = \kappa_1 + \kappa_2$. For this approach, we set ${w} = \chi$, $c_1 = \chi\kappa_1$, $c_2 = \chi\kappa_2$, and, in order to guarantee a faster response, we adopt $\kappa_1 = \kappa_2 = 2.05$~\cite{art:clerc2002particle}. For the velocity update expression, we consider $\{r_1, r_2\} \sim \mathcal{U}(0,1)$, where $\mathcal{U}(a,b)$ denotes a continuous uniform distribution with interval $[a,b]$. We express the particle's positions as vectors, ${\vec x}_i(t) \in \mathbb{C}^{1 \times 2N} = [|\alpha_1|, ..., |\alpha_K|, ..., |\alpha_N|, \theta_1, ..., \theta_N]$. The first subset of each particle $\{|\alpha_1|, ..., |\alpha_K|\}$ represents the amplitudes of the active coefficients of HR-RIS, the second subset $\{|\alpha_{K+1}|, ..., |\alpha_N|\}$ = \textbf{1} represents the remaining amplitudes for the passive elements of HR-RIS,  and for the last subset, $\{|\theta_1|, ..., |\theta_N|\}$ represents the phase shifts.

By noticing that an equivalent form of the optimization problem in~\eqref{opt:Cs_explicit}, $CF$ can be expressed as 
\begin{align}
CF({\vec x}) =& - \frac{1 + {\vec w}^\mathrm{H}  \boldsymbol{\mathrm{L}} {\vec w} }{1 + {\vec w}^\mathrm{H}  \boldsymbol{\mathrm{E}} {\vec w} } + P_c \left(\sum_{n = 1}^{K} |{\vec x}_n|^2 \xi_n - P_\mathrm{max}\right), \label{opt:CF}
\end{align} 
where $P_c$ is the punishment constraint. The Algorithm 1 starts by randomly generating $n_\mathrm{P}$ particles, where each particle is limited to ${\vec x}_\mathrm{min}$ and ${\vec x}_\mathrm{max}$. We set the initial personal best position to ${\vec p}_1(1) = {\vec x}_1(1)$ and the global best position to ${\vec p}_\mathrm{g}(1) = {\vec x}_n$, such that $CF({\vec x}_n)< CF({\vec x}_i), \forall i = 1,\cdots, n_\mathrm{P}, i \neq n $. Then, steps 3--16 are executed in order to find the optimal local solution, ${\vec p}_\mathrm{g}(m_\mathrm{I})$. Notice that it guarantees convergence due to the fact that for each iteration, all particles are moving towards the personal and global optimum solution.




\section{Numerical Results and Discussions}\label{sec:results}
In this section, we evaluate the secrecy performance of the proposed system through simulations of illustrative cases. According to real-world LOS channel measurements~\cite{art:MIMO_mmWave_LIS}, $a_\mathrm{LOS}, b_\mathrm{LOS}$ and $\mu_\mathrm{LOS}$ are, respectively, given by 61.4, 2, and 5.8~dB, and for NLOS channel measurements, $a_\mathrm{NLOS} = 72, b_\mathrm{NLOS}=2.92$, and $\mu_\mathrm{NLOS} = 8.7$~dB. For this system, we set the bandwidth to ${B} = 251.2$~MHz, and the noise figure to ${NF} = 10$~dB, and the noise power is set to $\sigma^2 = -174 + 10\log({B}) + {NF} \approx -80$~dBm. Unless otherwise specified, the transmit power at Alice is set to $P_T = 20$~dBm, the number of reflecting elements in the HR-RIS to $N$ = 40, the number of active elements at HR-RIS to $K = 2$,  $N_\mathrm{A} = 4, N_\mathrm{B} = 2$ and $N_\mathrm{E} = 2$, $~\epsilon = 0.1$, and the total number of resolvable paths for each Saleh-Valenzuela channel are $L_\mathrm{AR} = L_\mathrm{RB}=L_\mathrm{AB}=L_\mathrm{RE}= L_\mathrm{AE} = 3$. For Algorithm~1 parameters, we consider a population size of~$\textit{nPop} = 20$, the maximum number of iterations to $\textit{mIt} = 30$, and the punishment constraint to $P_c = 10^3$. We consider a two-dimensional topology network, where Alice, HR-RIS, Bob and Eve are respectively located at (0,0), (80,2), (90,0) and (100,0). Furthermore,  the simulation results are averaged over 1000 independent channel realizations. For comparison, we evaluate the secrecy capacity of the following benchmark schemes: (i) Case without RIS: optimization of transmit beamforming only; (ii) Case with optimal beamforming via~\eqref{eq:wOP}, and optimal phases considering a passive RIS: an AO-based approach is considered, where the amplitudes of RIS are set to 1,  and the phases are optimized based on Algorithm 1; (iii) Case with optimal beamforming as in~\eqref{eq:wOP} and optimal HR-RIS: a joint optimization between the transmit beamforming $\boldsymbol{\mathrm{w}}$, amplitudes and phase components of the HR-RIS is proposed based on~\eqref{eq:wOP} and Algorithm 1.

\begin{figure}[bt]
    \centering
      \vspace{-0.5cm}
    \includegraphics[width=9cm]{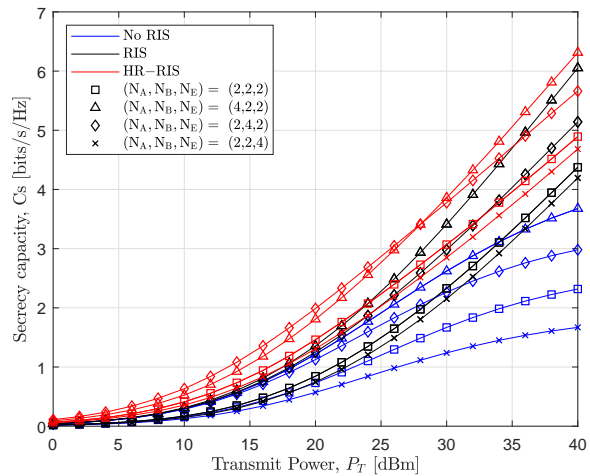}
    \caption{Secrecy capacity versus transmit power, considering different number of antenna configurations for $N_\mathrm{A}, N_\mathrm{B}$, and $~N_\mathrm{E}$. The corresponding parameters are: $\sigma_{\Delta} = 0.1, N = 40$. For the HR-RIS parameters, we consider $K = 2, P_\mathrm{max}$ = 10 dBm.}
    \label{fig:CsPt}
\end{figure}
Fig.~\ref{fig:CsPt} shows the secrecy capacity versus the transmit power at Alice, $P_T$, assuming different configurations of antenna elements at Alice, Bob and Eve. In addition, we set $\epsilon = 0.1$ and the reflecting coefficients at both passive RIS and HR-RIS to $N = 40$. For the HR-RIS parameters, we fix the number of active elements to $K = 2$, and the maximum power budget for the active elements to $P_\mathrm{max}$ = 10 dBm. The secrecy capacity for the HR-RIS scheme outperforms their counterparts, passive RIS and without RIS. Furthermore, in the HR-RIS scheme, the secrecy capacity performance improves as more antennas are employed at Bob node, considering a low-to-medium transmit power regime. Otherwise, at high transmit power regime, a larger number of antennas at the Alice leads to the best performance in terms of secrecy capacity. We can also observe that, even for the HR-RIS scheme in the worst case scenario (i.e., for $(N_\mathrm{A}, N_\mathrm{B}$, $N_\mathrm{E}) = (2,2,4)$), it is shown that HR-RIS allows higher secrecy rates than a passive RIS case with equal number of antennas at each node (i.e., for $(N_\mathrm{A}, N_\mathrm{B}$, $N_\mathrm{E}) = (2,2,2)$).

\begin{figure}[bt]
    \centering
    \vspace{-0.5cm}
    \includegraphics[width=9cm]{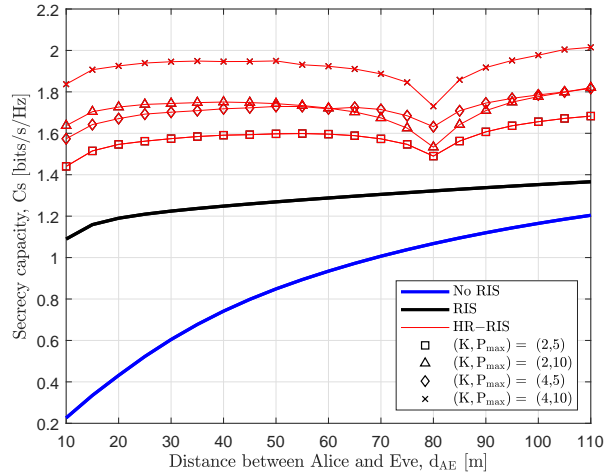}
    \caption{Secrecy capacity versus distance between Alice and Eve, considering $ N = 40, \sigma_{\Delta} = 0.1$, and $(N_\mathrm{A}, N_\mathrm{B}$, $N_\mathrm{E}) = (4,2,2)$. For the HR-RIS parameters, we consider $K = \{2, 4\}$ and $P_\mathrm{max} = \{5, 10\}$~dBm.}
    \label{fig:Csdae}
    \vspace{-0.6cm}
\end{figure}
Fig.~\ref{fig:Csdae} shows the secrecy capacity as a function of the distance between Alice and Eve, $d_\mathrm{AE}$, considering different configurations of $K$ and $P_\mathrm{max}$. Note that the system secrecy capacity performance can be improved significantly with the use of the HR-RIS scheme, even when the eavesdropper is closer to Alice. This can be explained by the fact that, from the closed-form solution in~\eqref{eq:wOP}, the transmit beamforming is designed to add the transmitted signal constructively at Bob while adding destructively at Eve's node. For the HR-RIS scheme, note that the secrecy capacity decreases when the eavesdropper node approaches the HR-RIS. This behavior can be caused because, in this case, the channel condition for the HR-RIS$\rightarrow$Eve is dramatically better than HR-RIS$\rightarrow$Bob counterpart. Also, even with the optimized solution for the active and passive coefficients, a limited and fixed active elements are not sufficient to overcome this issue. For a higher $P_\mathrm{max}$, note that a higher secrecy capacity performance is attained. However, when eavesdropper's positions are closer to the HR-RIS, the secrecy capacity is highly compromised. All in all, the HR-RIS system outperforms the other benchmark schemes in terms of the secrecy capacity.

\begin{figure}[bt]
    \centering
      \vspace{-0.8cm}
    \includegraphics[width=9cm]{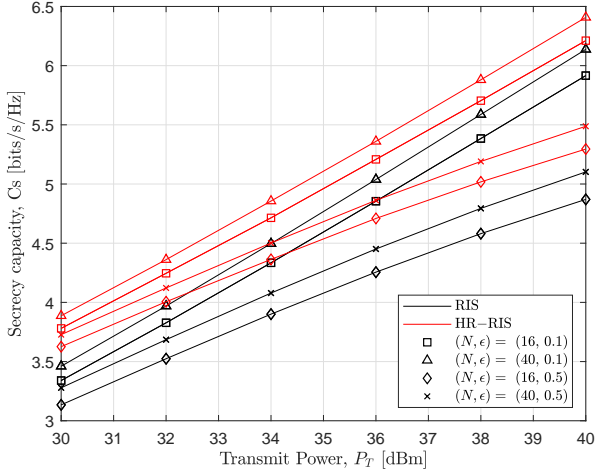}
    \caption{Secrecy capacity versus transmit power, for $N = \{16, 40\}$ and $\sigma_{\Delta} = \{0.1, 0.5\}$, considering $(N_\mathrm{A}, N_\mathrm{B}$, $N_\mathrm{E}) = (4,2,2)$. For the HR-RIS parameters, we consider $K = 2, P_\mathrm{max}$ = 10 dBm.}
    \label{fig:CsPtN}
     \vspace{-0.4cm}
\end{figure}
Fig.~\ref{fig:CsPtN} illustrates the secrecy capacity versus the transmit power at Alice, $P_T$, for different numbers of reflecting coefficients and different normalized variance of the error component, $\epsilon$, considering the passive RIS and HR-RIS schemes. For a high transmit power regime, by assuming the same CSI's eavesdropper accuracy, the proposed HR-RIS scheme with $N = 14$ passive reflecting coefficients and $K = 2$ active relay coefficients outperforms the conventional RIS with $N = 40$ reflecting coefficients' counterpart. It is also observed that a larger normalized error variance leads to an inaccurate joint transmit beamforming design at Alice and reflecting coefficients at RIS, resulting in a lower system secrecy capacity. 
\section{Conclusion}
In this work, we investigated the secrecy capacity maximization of an HR-RIS-assisted mmWave MIMO system, in the presence of a passive multi-antenna eavesdropper, with imperfect CSI from eavesdropping links. We proposed an AO-based algorithm by using the closed-form solution for the transmit beamformer and PSO. For comparison purposes, we considered two benchmark schemes referred to as $\mathrm{Op_{RIS}}$ and $\mathrm{No~RIS}$ schemes, which consider an optimized passive RIS scheme and a system with no RIS, respectively. From our results, the secrecy capacity for the proposed HR-RIS scheme showed to outperform both $\mathrm{Op_{RIS}}$ and $\mathrm{No~RIS}$ counterparts. 

\section*{Acknowledgements}

This work has been supported in part by Academy of Finland, 6G Flagship program (Grant 346208), FAITH Project (Grant 334280), as well as Infotech of Oulu Graduate School.


\balance

\bibliographystyle{IEEEtran}

\bibliography{main_Nhancomments}

\end{document}